\documentstyle[aps,prb,preprint]{revtex}
\begin{document}
\draft
\title{Effect of Proximity Coupling of Chains and Planes on the Penetration
Depth Anisotropy in YBa$_2$Cu$_3$O$_7$}
\author{W. A. Atkinson and J. P. Carbotte}
\address{Department of Physics and Astronomy, McMaster University, \\
Hamilton, Ontario, Canada L8S 4M1}
\date{\today}
\maketitle
\begin{abstract}
We calculate the penetration depth $\lambda$ in the $a$, $b$ and $c$
directions
for a simple model of YBa$_2$Cu$_3$O$_7$.  In this model there
are two layers---representing a CuO$_2$ plane and a CuO chain---per
unit cell.  There is
a BCS--like pairing (both $s$ wave and $d$ wave are considered) interaction
localised in the CuO$_2$ planes.
The CuO chains become superconducting at temperatures lower than
 $T_c$ because of
their proximity to the planes, and there is an induced gap in the
chains.  Since the temperature dependence of the penetration depth
in the $b$ direction (along the chains) is sensitive to the size of
the induced gap, the difference between the shapes of the penetration
depth curves in the $a$ and $b$ directions reveals a great deal about
the nature of the condensate in the chains.  We find that in our
proximity model there are always regions of the chain Fermi surface
on which the induced gap is much smaller than $T_c$, so that the
temperature dependence of $\lambda_b$ is always different than that
of $\lambda_a$.  Experimental observations of the
of the $ab$ anisotropy show nearly identical temperature dependences.
The main result of our paper, then, is that a simple proximity model
in which the pairing interaction is localized to the planes, and
the planes are coherently coupled to the chains cannot
account for the superfluid on the chains.
\end{abstract}
\narrowtext
\section{Introduction}
It is widely believed that the source of the pairing interaction which
is responsible for the superconducting trasition in the high $T_c$
cuprates lies in the CuO$_2$ planes, which are common to all of the
cuprates.    Many models which attempt to explain high $T_c$
superconductivity are two--dimensional, which is a reflection
of the assumption that the only active pieces of the crystal
are the CuO$_2$ planes and that the remaining ions act as
placeholders or as charge reservoirs.  In some materials, however,
there are additional layers whose behaviour is not clear.
In Bi$_2$Sr$_2$CaCu$_2$O$_8$, for example, it has been suggested
that the BiO layer plays the role of a normal metal in
close proximity to a superconducting material.\cite{Abrikosov}

The only materials in which there is clear evidence that the CuO$_2$
planes are not the only active portion of the unit cell are
YBa$_2$Cu$_3$O$_7$ (Y-123), YBa$_2$Cu$_4$O$_8$ (Y-124) and their
close relatives.
In these materials there are quasi--one dimensional CuO chain structures.
Experiments measuring the d.c.\ resistivity,\cite{Friedmann,Taillefer}
the infrared and optical conductivity,\cite{Basov} and the
penetration depth in untwinned crystals\cite{Bonn} and ceramics\cite{Tallon}
 have found
large anisotropies between the $a$ direction (in--plane, perpendicular
to the chains) and the $b$ direction (in--plane, parallel to the
chains) which suggest that substantial currents are carried along
the chains in both the normal and superconducting state.

In the superconducting state, the source of the condensate on the
chains in unclear.  One possibility is that the pairing interaction
is localized to the CuO$_2$ planes, but that the chains  become
superconducting by a proximity effect.  In the proximity effect,
an intrinsically normal metal which is in close contact with a
superconductor becomes superconducting near the junction as a
result of pair tunneling through the junction.  The size of the
induced gap in the normal metal depends on the strength of the
coupling across the junction.
Y-123 and Y-124 are good candidates for proximity effect models
because they have the least anisotropy between the in--plane
and $c$ axis transport properties of the cuprate superconductors, and
should therefore have a relatively large
coupling between the chains and planes.

Proximity effect models have been studied in the context of
high $T_c$ materials for a number of years.  The most common point
of view is that the unusual properities of the cuprates can
be explained by an isotropic BCS pairing interaction
which is contained in one of the planes.\cite{Abrikosov,Tachiki,Buzdin,%
Bulaevskii,Simonov,KlemmI,Kresin}  The idea behind most of the work is
that although the pairing interaction may be inherently isotropic,
the strongly anisotropic band structure leads to a gap structure
which may account for the unusual superconducting properties
of the cuprates.  The current authors have taken a different
point of view in recent work on proximity effect models.\cite{AtkinsonI%
,AtkinsonII,AtkinsonIII}  We have assumed that the pairing interaction
in the planes is intrinsically $d$ wave and then attempted to assess
the influence of coupling to the chains.
Closely related to the proximity models are the S/S$^\prime$
multilayer models in which there are two (or more) different
superconducting layers in the unit cell.  There have been
detailed examinations of the roles played by interplane
and intraplane pairing \cite{KlemmII,Schneider,Kettemann}
and a few quantitative calculations of physical properties,
\cite{Buzdin,KlemmII,Schneider} but these models have been
less thoroughly
explored than the proximity model because of their relative complexity.

In this article we address the issue of whether a proximity
model can account for the condensate on the chains in the YBaCuO
compounds.
We do this by calculating the penetration depth $\lambda$ for a simple
$s$ or $d$ wave model in the $a$, $b$ and $c$ directions.  In particular,
we are interested in comparing the temperature dependences of
$\lambda_a$ and $\lambda_b$ with experiment.\cite{Bonn}
In our model, the unit cell consists of a CuO$_2$ plane layer and
a CuO chain layer.  The CuO$_2$ planes contain the pairing interaction
and are coupled to the CuO chains through coherent single electron
tunneling so that there is an induced gap in the chains.
Calculations of the penetration depth in a similar
model have been made before,\cite{KlemmI,AtkinsonII} although
the intrinsically normal layers were planes and not chains,
and the emphasis was on the anisotropy between $\lambda_{ab}$ and
$\lambda_c$.

In Sec.\ \ref{Ham} we introduce our model Hamiltonian and find
the single particle Green's functions which we will need for the
penetration depth.  In Sec.\ \ref{depth} we derive an expression
for the penetration depth which is suitable for a two band,
tight binding model.  The calculation differs slightly from one
we made previously.\cite{AtkinsonII}  In Sec.\ \ref{res} we
discuss the results of numerical calculations of the penetration
depth, and in Sec.\ \ref{con} we broaden the scope to a discussion of
the nature of the condensate on the chains.

\section{Hamiltonian}
\label{Ham}
The goal of this section is to introduce our model for YBCO,
and to find the single particle Green's functions necessary for
the calculation of the penetration depth in Sec.\ \ref{depth}.
We begin with a Hamiltonian which describes a system with two layers
per unit cell.  Adjacent layers are separated by a distance $d/2$.
The first layer represents a CuO$_2$ plane and it
contains a BCS--like pairing interaction.  The second layer represents
a CuO chain.  It has a one dimensional dispersion and is intrinsically
normal.  The chains are superconducting, however, because of their
coupling to the planes through single electron tunneling.  The
Hamiltonian, expressed in the Nambu formalism, is:
\begin{equation}
  \label{101}
  {\bf H} - {\bf N}\mu = \sum_{\bf k} {\bf C}^\dagger ({\bf k})Q({\bf k})
  {\bf C}({\bf k}) + \mbox{const.},
\end{equation}
where
\begin{equation}
  \label{102}
  {\bf C} ({\bf k}) = \left [ \begin{array}{c}
  {\bf c}_{1 {\bf k} \uparrow} \\
  {\bf c}^\dagger_{1 -{\bf k} \downarrow} \\
  {\bf c}_{2 {\bf k} \uparrow} \\
  {\bf c}^\dagger_{2 -{\bf k} \downarrow}
  \end{array} \right ]
\end{equation}
and
\begin{equation}
  \label{103}
  Q =  \left [ \begin{array}{cccc}
  \xi_1({\bf k}) & -\Delta_{\bf k} & t({\bf k}) & 0 \\
  -\Delta^\ast_{\bf k} & -\xi_1(-{\bf k}) & 0 & -t^\ast(-{\bf k}) \\
  t^\ast({\bf k}) & 0 &  \xi_2({\bf k}) & 0 \\
  0 &  -t(-{\bf k}) & 0 & -\xi_2(-{\bf k})
  \end{array} \right ] .
\end{equation}
This Hamiltonian has been discussed at length elsewhere \cite{Abrikosov,%
AtkinsonI,AtkinsonIII} and we only describe it briefly here.  The dispersions
$\xi_1$ and $\xi_2$ are for the plane and chain layers respectively.
We assume tight binding dispersions so that $\xi_1 = -2\sigma_1
[\cos(k_x a) + \cos(k_y b)] - \mu_1$ and $\xi_2 = -2\sigma_2 \cos(k_y b)
-\mu_2$, where $a$ and $b$ are the lattice constants in the planes.
In optimally doped Y-123 crystals, $a$ and $b$ differ by $\sim 1 \%$.
For the numerical calculations done in this work, we take $a = b$.
We take $\sigma_1 = 100 \mbox{ meV}$ so that the full
bandwidth of the CuO$_2$ plane is 0.8 eV.
For the chain layer we take $\sigma_2 = 80 \mbox{ meV}$ so that
$\lambda_a^2(T=0)/\lambda_b^2(T=0)
\sim 2.5$, as seen experimentally by Basov {\em et al.} \cite{Basov}
The chemical potentials are $\mu_1 = -80 \mbox{ meV}$
and $\mu_2 = 40 \mbox{ meV}$, which yields a Fermi surface that
is in qualitative agreement with band structure calculations.
\cite{Pickett,Yu}  We have absorbed an arbitrary band offset into the
chemical potentials so that $\mu_1 \neq \mu_2$.  The chains and planes
are coupled by the matrix element $t(k_z) = - t_0 \cos(k_z d/2)$,
where $d/2$ is the distance between the chains and planes.
The chain--plane coupling affects the penetration depth in two
ways.  First, $t(k_z)$ determines the $c$--axis transport properties.
In a previous paper \cite{AtkinsonII} we have shown that the ratio
$\lambda_c(T=0)/\lambda_{ab}(T=0)$ varies inversely with $t_0$.
Second, $t(k_z)$ determines the size of the superconducting gap
induced on the chains, which is reflected in the temperature dependence
of  $\lambda_b$.  In this article we choose $t_0 = 50 \mbox{ meV}$
which yields $\lambda_c^2(T=0)/\lambda_a^2(T=0) \sim 100$, which is
in rough agreement with experimental observations.\cite{BasovII,%
Anlage,Schutzmann,Homes}

The final feature of our Hamiltonian is that there is a
pairing interaction in the plane which drives the superconducting
transition.  As we have mentioned above, the chains also become
superconducting at $T_c$ through their coupling
to the planes.  The pairing interaction in the planes has the form
$V_{{\bf k} {\bf k}^\prime} = V\eta_{\bf k}\eta_{{\bf k} ^\prime}$ with
$\eta_{\bf k} = 1$ for an {\em s}-wave superconductor and
$\eta_{\bf k} = \cos(k_x) - \cos(k_y)$ for a d$_{x^2-y^2}$ superconductor.
Since the pairing interaction is separable, the order parameter
\begin{equation}
\label{104}
\Delta_{\bf k} \equiv \frac{1}{\Omega} \sum_{{\bf k} ^\prime}
  V_{{\bf k k}^\prime} \langle {\bf c}_{1 -{\bf k}^\prime \downarrow}
  {\bf c}_{1 {\bf k}^\prime \uparrow} \rangle,
\end{equation}
(where $\Omega$ is the volume of the crystal)
can be written $\Delta_{\bf k} = \Delta_0 \eta_{\bf k}$.

Diagonalization of the Hamiltonian leads to four energy bands $E_1 = E_+$,
$E_2 = E_-$, $E_3 = -E-$, $E_4 = -E+$ with
\begin{eqnarray}
  \label{105}
  E_\pm ^2 & = & \frac{\xi_1^2 + \xi_2^2 + \Delta_{\bf k}^2}{2} + t^2
  \nonumber \\
  & \pm &\sqrt{ \left [ \frac{\xi_1^2 - \xi_2^2 + \Delta_{\bf k}^2}{2}
  \right ]^2 + t^2 [ (\xi_1 + \xi_2)^2 + \Delta_{\bf k}^2] } .
\end{eqnarray}
In Fig.\ \ref{f0} we show the Fermi surface for a range of $k_z$ values
between 0 and $\pi/d$.  The Fermi surface consists of two surfaces on
which $E_-$ vanishes in the normal state.  The two surfaces are given by
the two solutions to $\xi_1({\bf k})\xi_2({\bf k})=t(k_z)^2$.
When $k_z=\pi/d$, $t(k_z) = 0$ and
the two pieces of Fermi surface are those of
the isolated chain and plane subsystems.  When $t(k_z) \ne 0$,
the chain and plane states form hybrid bands whose energies are
given by
\[
  \epsilon_\pm = \frac{\xi_1+\xi_2}{2} \pm \sqrt{\left(
\frac{\xi_1-\xi_2}{2}\right) ^2 + t^2},
\]
in the normal state.
The shift in the band energy due the chain--plane coupling is clearly
dependent on the relative sizes of $t^2$ and $(\xi_1 - \xi_2)^2$.  The
effect of the chain--plane coupling on the Fermi surfaces shown in
Fig.\ \ref{f0} is largest in the neighbourhood of the Fermi surface
crossing at $\xi_1 = \xi_2 =0$.

The quasiparticle operators in the diagonalized representation are
\begin{equation}
  \label{105b}
  {\bf \hat{C}}_i({\bf k}) = \sum_{j=1}^{4} U^\dagger_{ij} ({\bf k})
  {\bf C}_j ({\bf k}),
\end{equation}
where $U ({\bf k})$ is the $4 \times 4$ matrix which diagonalises $Q$:
$U = \left [ U_1\,U_2\,U_3\,U_4 \right ]$ with
\begin{eqnarray}
  \label{106}
  U_j& = & \frac{1}{\sqrt{C}} \left [\begin{array}{c}
               (E_j - \xi_2) A \\
	       -(E_j + \xi_2) B \\
	       t A \\
	       t B \\
	       \end{array} \right ]
\end{eqnarray}
\[
  A  =  t^2 - (\Delta_{\bf k} + E_j + \xi_1)(E_j + \xi_2)
\]
\[
  B  =  t^2 - (\Delta_{\bf k}^\ast + E_j - \xi_1)(E_j - \xi_2)
\]
\[
  C =  A^2 [t^2 + (E_j - \xi_2)^2] + B^2 [t^2 + (E_j + \xi_2)^2].
\]

Now that we have diagonalized the Hamiltonian, we can find the
single particle Green's functions which we will require in the
followin section.  Defining the temperature
Green's function $G({\bf k};\tau)_{ij} =
-(1/\hbar)\langle {\bf T} {\bf C}_i({\bf k};-i\tau) {\bf C}^\dagger_j
({\bf k};0)
\rangle $, where ${\bf T}$ is the fermion time--ordered product, we have
\begin{equation}
  \label{107}
  G({\bf k};i\zeta_l)_{ij} = \sum_{m=1}^4 \frac{U_{im}({\bf k})
  U_{mj}^\dagger({\bf k})}{i\hbar\zeta_l - E_m({\bf k})} ,
\end{equation}
where
\[
  G({\bf k};i\zeta_l)_{ij} = \int_0^{\hbar\beta} d\tau \, e^{i\zeta_l \tau}
  G({\bf k};\tau)_{ij} ,
\]
$\zeta_l = (2l+1)\pi/\hbar\beta$ are the fermion Matsubara frequencies and
$\beta = 1/k_BT$, where $k_B$ is the Boltzmann constant and $T$ is
temperature.  In our calculation of the penetration
depth we will need
\begin{equation}
  \label{107.5}
  G({\bf k};\tau = 0^-)_{ij} = \frac{1}{\hbar}\sum_{l=1}^4 U_{il}({\bf k})
   f[E_l({\bf k})] U_{lj}^\dagger({\bf k}),
\end{equation}
where $f(x) = 1/[1+\exp(\beta x)]$.


We finish this section with a brief discussion of our usage of the word
``gap''.  In the model presented above, there is only one order parameter,
$\Delta_{\bf k}$, and it describes the condensate in the CuO$_2$ planes.
For a separable potential, $\Delta_{\bf k}$ has the symmetry of the
pairing interaction.  In a multiband material, however, $\Delta_{\bf k}$
is not simply related to the pair wavefunction.  For example, the
anomalous Green's function (which is essentially the pair wavefunction)
in the CuO$_2$ plane is
\[
  G({\bf k};\omega)_{12} = -\frac{\Delta_{\bf k}(\omega^2-\xi_2^2)}
{(\omega^2-E_+^2)(\omega^2-E_-^2)}.
\]
Notice that the symmetry of $G_{12}$ is {\em not} the same as the symmetry
of $\Delta_{\bf k}$.  For this reason, the term ``gap'' is kept distinct
from the term ``order parameter'', which refers to $\Delta_{\bf k}$.
Perhaps the most useful working definition of ``gap'' is that it is the
value of $E_-$ on the Fermi surface.  Clearly, by this definition, the gap
is ${\bf k}$--dependent.  In
regions of the Brillouin zone where a section of Fermi surface has
predominantly chain (or plane) character, the gap can be associated with
the chains (or planes).
It is wrong to think of the pairs being localised to the chains or planes,
however; the pairing amplitude between an electron in the chains and
an electron in the planes (measured by $G_{14}$ and $G_{23}$) is nonzero.
In fact the picture of
a gap belonging to a plane or a chain breaks down in regions of the Brillouin
zone where the Fermi surface is a strong hybridization of the chain and
plane bands.


\section{Penetration Depth}
\label{depth}
The penetration depth is found using an approach which is suitable
for the tight binding limit.  This approach is slightly different
than that of our previous work,\cite{AtkinsonII} although it yields
quantitative results which are nearly identical.  As we shall
see, however, the current method, which is based on one used by Hirsch
and Marsiglio for a one band tight binding model,\cite{Hirsch} is more
satisfying from a physical point of view.

We begin by writing
out the current operator ${\bf j_0}$ in the absence of a magnetic field:%
\cite{Mahan}
\[
  {\bf j_0}({\bf q}=0) \equiv \int d^3{\bf r \, j}({\bf r})
\]
\begin{equation}
  \label{108}
  = \frac{ie}{\hbar} [{\bf H_0^n},{\bf P}],
\end{equation}
where
\begin{equation}
  \label{109}
  {\bf P} \sim \sum_{i,{\bf R_i}} {\bf R_i} {\bf c}_i^\dagger({\bf R_i})
 {\bf c}_i({\bf R_i}),
\end{equation}
is the polarization vector
and ${\bf H^n}$ is the Hamiltonian in the normal state.
The operator ${\bf c}^\dagger_i({\bf R_i})$ creates an electron in the
Wannier state located at the sublattice point ${\bf R_i}$.
The set of points $\{ {\bf R_1} \}$ refers to the plane sublattice while
$\{ {\bf R_2} \}$ refers to the chain sublattice.
The Wannier representation is connected to the ${\bf k}$--space
representation by
\begin{equation}
  \label{110}
  {\bf c}_i({\bf R_i}) = \frac{1}{\sqrt{N}} \sum_{\bf k} e^{i{\bf k
  \cdot R_i}}{\bf c}_i({\bf k}),
\end{equation}
where $N$ is the number of lattice sites.

In the normal state the Hamiltonian, Eq.\ (\ref{101}), can be written
in the Wannier representation as:
\begin{eqnarray}
  \label{111}
  {\bf H^n_0} &=& - \sum_{i=1}^2 \sum_{{\bf R_i},{\bf r_i}}
  \sigma_i
  {\bf c}^\dagger_i({\bf R_i}+{\bf r_i}){\bf c}_i({\bf R_i})\nonumber \\
  &-& \frac{t_0}{2}\sum_{\bf R_1,R_2} \left \{
  {\bf c}^\dagger_1({\bf R}_1)
  {\bf c}_2({\bf R_2}) \left [ \delta_{{\bf R}_1+{\bf \hat{z}}d/2,
  {\bf R_2}} \right. \right. \nonumber \\
  &+& \left. \left. \delta_{{\bf R}_1-{\bf \hat{z}}d/2, {\bf R_2}} \right ]
  + \mbox{H.c.} \right \}
\end{eqnarray}
The vector ${\bf r_i}$ is the displacement to the nearest neighbours of
${\bf R_i}$ within the plane, ${\bf \hat{z}}$ is the unit vector in the
{\em z}-direction, and H.c. indicates the Hermitian conjugate.  This
Hamiltonian describes nearest neighbour hopping both within and
between the chains and planes.
Substituting Eqs.\ (\ref{109}) and (\ref{111}) into Eq.\ (\ref{108})
we get
\begin{eqnarray}
  \label{112}
  {\bf j_0} &=& \frac{ie}{\hbar} \sum_{i,{\bf R_i},{\bf r_i}}
  \sigma_i \, {\bf r_i}
  {\bf c}^\dagger_i({\bf R_i}+{\bf r_i}){\bf c}_i({\bf R_i})\nonumber \\
  &-& \frac{iet_0 d}{4\hbar} \sum_{\bf R_1} \hat{\bf z} \left[
  {\bf c}^\dagger_1({\bf R}_1) {\bf c}_2({\bf R}_1+{\bf \hat{z}}d/2)
  \right.  \nonumber \\
  &-& \left. {\bf c}^\dagger_1({\bf R}_1) {\bf c}_2({\bf R}_1-
  {\bf \hat{z}}d/2) - \mbox{H.c.} \right ] .
\end{eqnarray}

In the presence of a finite magnetic vector potential ${\bf A}({\bf r})$,
the tight binding Wannier states are modified by a phase so that
\begin{equation}
  \label{113}
  {\bf c}_i({\bf R_i}) \rightarrow {\bf c}_i({\bf R_i})
  \exp \left [ -\frac{ie}{\hbar c}{\bf R_i \cdot A}({\bf R_i}) \right ],
\end{equation}
where $c$ is the speed of light.
The assumption is made that the vector potential is slowly varying over
the length scale of the crystal lattice, and we will make use of the
fact that ${\bf A}({\bf q})$ is strongly peaked about ${\bf q}= 0$
throughout this section.
To linear order in ${\bf A}$, then, Eq.\ (\ref{111}) becomes
\begin{equation}
  \label{114}
  {\bf H^n} = {\bf H^n_0} - \frac{1}{c\Omega}{\bf j_0}\cdot{\bf A}
  ({\bf q=0})
\end{equation}
and Eq.\ (\ref{112}) becomes
\begin{eqnarray}
  \label{115}
  {\bf j} &=& {\bf j_0} - \frac{e^2 }{\hbar^2 c} \sum_{i,{\bf R_i,r_i}}
  \sigma_i\, {\bf r_i}[{\bf r_i \cdot A}({\bf R_i})] {\bf c}^\dagger_i
  ({\bf R_i}+{\bf r_i}) {\bf c}_i({\bf R_i}) \nonumber \\
  & - & \frac{e^2 t_0 d^2}{8 \hbar^2 c}
  \sum_{\bf R_1} {\bf \hat{z}}[{\bf \hat{z}\cdot A}({\bf R_i})] \left [
  {\bf c}^\dagger_1({\bf R_1}) {\bf c}_2({\bf R_1}+{\bf \hat{z}}d/2)
  \right. \nonumber \\
  &+& \left.  {\bf c}^\dagger_1({\bf R_1}) {\bf c}_2({\bf R_1}-
  {\bf \hat{z}}d/2) + \mbox{H.c.} \right ].
\end{eqnarray}
In the presence of a magnetic field, the observable current is given
by $\langle {\bf j} \rangle$, and not $\langle {\bf j_0} \rangle$.
We can rewrite Eq.\ (\ref{115}) in a ${\bf k}$--space
representation using Eq.\ (\ref{110}):
\begin{eqnarray}
  \label{115.5}
  {\bf j}({\bf q}=0) &=& \frac{e}{\hbar} \sum_{\bf k} \left [ {\bf C}^\dagger
  ({\bf k}) \vec{\gamma}_{\bf k} {\bf C}({\bf k}) \right. \nonumber \\
  &-& \left. \frac{e}{\hbar c \Omega}
  {\bf C}^\dagger ({\bf k}) \left[ \tensor{\gamma}^\prime_{\bf k}
  \cdot {\bf A} ({\bf q} = 0)
  \right ]{\bf C}({\bf k}) \right ].
\end{eqnarray}
The vector $\vec{\gamma}_{\bf k}$ is a $4 \times 4$ matrix with
three spatial components.  It is essentially the Fermi velocity:
\[
  \vec{\gamma}_{{\bf k}ij} = (-1)^{i-1} \frac{\partial Q_{ij}}
  {\partial {\bf k}},
\]
where $Q_{ij}$ is the Hamiltonian matrix of Eq.\ (\ref{103}) and
the factor $(-1)^{i-1}$ comes from the fact that ${\bf C}_i$
annihilates electron states for $i=1,3$ and hole states for $i=2,4$.
The dyadic $\tensor{\gamma}^\prime_{\bf k}$ is also a $4 \times 4$
matrix and it is essentially the effective mass tensor:
\[
  \tensor{\gamma}^\prime_{{\bf k}ij} = \frac{\partial^2 Q_{ij}}
{\partial {\bf k} \partial {\bf k}}.
\]
The first term in Eq.\ (\ref{115.5}) is ${\bf j_0}({\bf q}=0)$ while
the second term contains the remaining two terms in Eq.\ (\ref{115}).

The current $\langle {\bf j} \rangle$ which
is generated by the applied magnetic field is given, to linear order in
${\bf A}$ by the Kubo formula:\cite{Rickayzen}
\[
  \langle {\bf j}(t) \rangle = \langle {\bf j}(t) \rangle_0
  + \frac{i}{\hbar} \int_{-\infty}^{t} dt^\prime \,
  \langle [-\frac{1}{c}{\bf j_0}(t^\prime) \cdot {\bf A}(t^\prime)
  ,{\bf j_0}(t)]
  \rangle_0.
\]
In the London limit, for the case of a static applied field,
this gives
\begin{equation}
  \label{116a}
  \langle {\bf j}({\bf r}) \rangle  = \langle {\bf j}({\bf r}) \rangle_0
  - \frac{1}{c \Omega} \sum_{\nu} G^j_{\mu\nu}
  (0,0;0) {\bf A}_\nu({\bf r}),
\end{equation}
where
\begin{eqnarray}
  G^j_{\mu\nu}({\bf q},{\bf q}^\prime,i\omega_l)&=&
  - \frac{1}{\hbar \Omega} \int_0
  ^{\hbar\beta} d\tau \, e^{i\omega_l \tau}  \nonumber \\
  & \times & \langle {\bf T} {\bf j_0}_\mu
  ({\bf q},-i\tau) {\bf j_0}_\nu({\bf q}^\prime,0) \rangle_0.
  \nonumber
\end{eqnarray}
$G^j$ is the current--current correlation function, ${\bf T}$ is the
boson time--ordered product, $\omega_l = 2l\pi/\hbar\beta$ are the
boson Matsubara frequencies and $\mu$ and $\nu$ refer to the
spatial components of ${\bf j_0}$.
The expectation values $\langle \, \rangle_0$ are taken with
respect to the zero field wavefunction.  It is straightforward
to evaluate $G^j$ in terms of the single particle Green's functions:
\begin{eqnarray}
  \label{117}
  G^j_{\mu\nu} (0,0;i\omega_l) &=&  \lim_{{\bf q} \rightarrow 0}
  \frac{e^2}{\beta \hbar^2 \Omega} \sum_{n,{\bf k}} \mbox{Tr} \left [
  G({\bf k};i\zeta_n - i\omega_l) \vec{\gamma}_{{\bf k}\mu}
  \right. \nonumber \\
  &\times& \left.
  G({\bf k}+{\bf q};i\zeta_n) \vec{\gamma}_{{\bf k}\nu} \right ].
\end{eqnarray}

The first term in Eq.\ (\ref{116a}) is the diamagnetic contribution to
the screening current.  Using Eqs.\ (\ref{107.5}) and (\ref{115.5})
we may evaluate this explicitly:
\begin{mathletters}
\begin{eqnarray}
  \label{118a}
  \langle {\bf j}({\bf r}) \rangle_{\mbox{\it dia}} & = & \frac{1}{\Omega}
  \sum_{\bf q} \langle {\bf j} ({\bf q}) \rangle_{\mbox{\it dia}}\,
  e^{i{\bf q \cdot r}} \nonumber \\
  & = &  \frac{e^2}{\Omega\hbar c} \sum_{\bf k}
  \mbox{Tr} \left [ G({\bf k};\tau=0^-) \tensor{\gamma}^\prime_{\bf k}
  \right ] \cdot {\bf A}({\bf r}) \nonumber \\
  & = & \frac{e^2}{\Omega \hbar^2 c} \sum_{\bf k} \sum_{i = 1}^4
    f[E_i({\bf k})]
  \hat{\gamma}^\prime_{{\bf k}ii}  \cdot {\bf A} ({\bf r})
\end{eqnarray}
where Tr is a trace over the components of the $4 \times 4$ matrix
contained in the square brackets, and
$\hat{\gamma}_{\bf k}^\prime = U^\dagger({\bf k})
\tensor{\gamma}_{\bf k}^\prime U({\bf k}) $.  In order to derive
Eq.\ (\ref{118a}), we have used the fact that $\langle {\bf j_0} \rangle_0
= 0$ and that ${\bf A}({\bf q})$ is peaked about ${\bf q} = 0$.
The second term in Eq.\ (\ref{116a}) is the paramagnetic contribution
to the screening current.  Evaluating Eq.\ (\ref{117}) explicitly
we have
\begin{eqnarray}
  \label{118b}
  \langle {\bf j} \rangle_{\mbox{\it para}} & = & -\frac{e^2}{\Omega
  \hbar^2 c}
  \sum_{i,j = 1}^4 \sum_{\bf k} \hat{\gamma}_{{\bf k}ij}
  [\hat{\gamma}_{{\bf k}ji} \cdot {\bf A}({\bf r})] \nonumber \\
  &\times & \left [ \delta_{i,j} \frac{\partial f(E_i)}{\partial E_i} +
  [1-\delta_{i,j}]\frac{f(E_i) - f(E_j)}
  {E_i - E_j} \right], \nonumber \\
\end{eqnarray}
\end{mathletters}
where $\hat{\gamma}_{\bf k} = U^\dagger({\bf k}) \vec{\gamma}_{\bf k}
U({\bf k})$. This expression for the paramagnetic current is the same
as in Ref.\ \onlinecite{AtkinsonII} where it was discussed at length.
We will only  repeat the points which are directly relevant to the current
work, and the interested reader is referred to our
earlier work.  The total current produced by the magnetic field is
\begin{eqnarray}
  \label{119}
  \langle {\bf j} \rangle & = & \langle {\bf j} \rangle_{\mbox{\it dia}}
  + \langle {\bf j} \rangle_{\mbox{\it para}} \nonumber \\
  & = & - {\bf K \cdot A}.
\end{eqnarray}
It is straightforward to show that
${\bf K}_{\mu\nu} = 0$ if $\mu \neq \nu$ (recall that $\mu$ and $\nu$
refer to spatial directions) so that the penetration depth
is given by
\begin{equation}
  \label{120}
  \frac{1}{\lambda_\mu^2} = {\bf K}_{\mu \mu}.
\end{equation}
This is the main result for this section.  In order to
plot $\lambda_\mu^{-2}$ as a function of temperature in Sec.\ \ref{res},
we must evaluate the integrals in Eqs.\ (\ref{118a}) and
(\ref{118b}) numerically.

We will finish this section with a few comments about Eqs.\ (\ref{118a})
and (\ref{118b}).  In the usual treatment of the penetration depth
the diamagnetic contribution to the screening current is
${\bf j}_{\mbox{\it dia}} = - ne^2{\bf A}/mc$,
which is independent of temperature.  The temperature dependence
of the penetration depth, then, comes from the paramagnectic
contribution to the screening currents which, for a one band free
electron metal with an isotropic gap, is
\begin{equation}
  \label{121}
  {\bf j}_{\mbox{\it para}} = -\frac{ne^2}{2\mu \hbar c} \int_{-\mu}^\infty
  d\epsilon \,[v^f]^2\, \frac{\partial f(E)}{\partial E},
\end{equation}
where $E = [\epsilon^2+\Delta^2]^{1/2}$, $v^f$ is the
Fermi velocity, and $\mu$ is the chemical potential.  The paramagnetic
term counts the number of thermal excitations (broken Cooper pairs)
which degrade the screening current.  At $T=0$ the paramagnetic term
vanishes, so that ${\bf j} = -ne^2{\bf A}/mc$.
When $\Delta = 0$ the paramagnetic term cancels the
diamagnetic term exactly so that ${\bf j} = 0$.  For systems
which are more complicated
than the free electon gas, it is common to make the approximation
$\left. {\bf j}_{\mbox{\it dia}} = -{\bf j}_{\mbox{\it para}} \right
|_{\Delta=0}$.  The approximation is exact at $T=T_c$ and, provided
the temperature dependence of the diamagnetic term is weak,
the approximation is a good one.  This is the approximation we made
in our previous discussion of the two layer model.\cite{AtkinsonII}
In the current work, however, we have treated ${\bf j}_{\mbox{\it dia}}$
in a fashion which is more consistent with the tight binding model,
so that while Eq.\ (\ref{118b}) is the same as we found previously,
Eq.\ (\ref{118a}) is different.  There is
little quantitative difference between the two approaches, however,
since both expressions for the diamagnetic current are weakly temperature
dependent and both cancel the paramagnetic current above $T_c$.

The most significant difference between
Eq.\ (\ref{118a}) and the usual expression for the penetration depth is the
interband term, which is proportional to $[f(E_i)-f(E_j)]/[E_i-E_j]$.
While the intraband term (which is proportional to $\partial f(E_i)
/\partial E_i$) counts the number of thermally broken pairs,
the interband term describes the degradation of the screening
currents by interband transitions.  The interband term does not
vanish at $T=0$ so that, unlike the single band case, there is a
finite paramagnetic contribution to the screening current.

\section{Results}
\label{res}
The question we are attempting to address in this article is
whether a proximity effect model can account for the experimentally
observed anisotropy in the temperature dependence of the
penetration depth in Y-123.
In this section we will present the results of numerical calculations of
the penetration depth for the model Hamiltonian introduced in Sec.\
\ref{Ham}.  We will compare these results to experiments and to related
calculations made with a two--plane proximity model (in which the
intrinsically normal layer is a two dimensional plane).
One of the main goals of this section is to emphasize
the difference between two--plane proximity models and chain--plane
models of the type studied here.

To begin with, we will discuss calculations of the penetration
depth in the two--plane proximity models.\cite{KlemmI,AtkinsonII}
One of the important features of the proximity model is that it
introduces low energy excitations into the superconducting spectrum.
The reason for this is that the induced gap in the intrinsically
normal plane is proportional to the strength of the chain--plane
coupling $t(k_z)$ (which vanishes at $k_z=\pi/d$) so that the gap
will have a nodal structure even if the pairing interaction has
isotropic $s$ wave symmetry.  The need for a gap structure with
nodes has been suggested, for example, by measurements\cite{Hardy}
of $\lambda_{ab}$ (the in--plane penetration depth)
in twinned single crystals of Y-123. The linear dependence of
$\lambda_{ab}(T)$ on $T$ at low temperatures is easily explained
by any gap structure with nodes.\cite{Annett}
While these measurements are commonly taken as support for $d$ wave
models,\cite{Prohammer} it has also been shown\cite{KlemmI,AtkinsonII}
that two--plane proximity models also result in linear low $T$ behaviour.
Since a central theme in much of the work on proximity models
\cite{Abrikosov,Tachiki,Buzdin,Bulaevskii,Simonov,Kresin,KlemmII}
is that the pairing interaction in the intrinsically
superconducting plane is $s$ wave, the low energy excitations in the
induced gap are an essential feature of the proximity models.

In Fig.\ \ref{f1}(a) we plot the penetration depth for our plane--chain
proximity model for the case of an $s$ wave gap.  We find that,
unlike the case of the two-plane model, the temperature dependences
of $\lambda_a$ and $\lambda_b$ are dramatically different.  The
most important difference is that the temperature dependence of
$\lambda_a$ is nearly identical to that of a single layer $s$
wave material with no chains, while $\lambda_b$ has a linear
low $T$ behaviour similar to that found in the two--plane proximity
models.\cite{AtkinsonII}
The factor of two difference between $\lambda_a(0)^{-2}$
and $\lambda_b(0)^{-2}$ comes from the screening currents carried
in the $b$ direction by the chains, and the linear $T$ dependence
in $\lambda_b^{-2}$ at low temperatures comes from the node in the
induced gap at $k_z=\pi/d$.  The fact that the low $T$ behaviour
of $\lambda_a^{-2}$ is exponential and not linear indicates that
pairs associated with $a$ axis screening currents have a finite gap
for all values of $k_z$.  We can understand this in more concrete
terms as follows:  In Sec.\ \ref{depth} we showed that the screening
current has two parts---a diamagnetic part which is roughly independent
of $T$ and a paramagnetic part which accounts for processes
(such as thermal pair breaking) which degrade the screening currents.
The temperature dependence of the penetration depth comes from
the paramagnetic screening current,
given in Eq.\ (\ref{118b}).  Despite its complicated appearance,
Eq.\ (\ref{118b})
has a simple physical interpretation.  The factors $\hat{\gamma}_{\bf k}$
are electron Fermi velocity {\em vectors},
while the two terms involving Fermi functions count the number
of thermally excited quasiparticles which participate in intraband ($i=j$)
or interband ($i \neq j$) paramagnetic processes.  When we
calculate the screening current in the $a$ direction, then, the
integrand in Eq.\ (\ref{118b}) is weighted by the square of
the Fermi velocity in the $a$ direction.  In Fig.\ \ref{f0}, we
can see that this is small both on segments of the Fermi surface
associated with the chains and on segments of the plane Fermi
surface which are distorted by the chains.  The most obvious consequence
of this is that the chains do not participate significantly
in carrying currents in the $a$ direction.
A more subtle result is that, even though there is a node in the
induced gap in the chains, it is not seen by electrons travelling in
the $a$ direction so that Cooper pairs which are part of the
$a$ axis screening current have a finite gap.  The onset of thermal
pair breaking, then, occurs at a much lower temperature in
the $b$ axis supercurrent than in the $a$ axis supercurrent.

In Fig.\ \ref{f1}(b) we plot the penetration depth for a
$d$ wave order parameter and find results which are similar to the
$s$ wave case:  $\lambda_a(T)$ is essentially the same as found in
single layer $d$ wave models
and $\lambda_b(T)$ resembles $\lambda_{ab}(T)$ found in the
two--plane proximity models.  As for the case of an $s$ wave gap,
the reason is that there are
a larger number of low energy excitations in the chains than in the
planes.  The $d$ wave gap in the planes has nodes along $k_x = \pm k_y$,
while the induced gap has nodes along $k_x = \pm k_y$ and $k_z = \pi/d$.

The large temperature dependence of the $ab$ anisotropy
seen in Figs.\ \ref{f1}(a) and (b) is difficult to
reconcile with measurements of $\lambda_a(T)$  and $\lambda_b(T)$
in untwinned crystals.  In these experiments \cite{Bonn}
$\lambda_a$ and $\lambda_b$ have a nearly identical
temperature dependence, although their absolute magnitude differs
by a factor of 1.5 at $T=0$.  In our model, the temperature dependence
of the anisotropy is a result of the fact that
Cooper pairs in the chains are more easily broken than Cooper
pairs in the planes.  Clearly, then, in a realistic model,
the density of low energy excitations in the
chains must be similar to that in the planes.  This is not a trivial
requirement.  It implies that both the nodal structure and the magnitude
of the gaps in the chains and planes be similar.  It is possible
to eliminate the nodes in the induced gap at $k_z = \pi/d$  by, for
example, making the ansatz that $t(k_z) = t_0$ (this would describe
a single bi--layer).  However, this is not sufficient to eliminate
the temperature dependence of the anisotropy.
For regions of the chain Fermi surface where $|\xi_1| \gg |t(k_z)|$, the
induced gap is
of the order\cite{AtkinsonII} $\Delta_{\bf k}t(k_z)^2/\xi_1^2$.
In Fig.\ \ref{f0}, the smallest induced gap occurs at the intersection
of the chain Fermi surface with the Brillouin zone boundary (at $k_x =
\pi/a$) at which $t_0^2/\xi_1({\bf k})^2 \sim 0.023$.  The onset of
thermal pair breaking in the chains, therefore, will occur at a much
lower temperature than in the planes.

The penetration depth in the $c$ direction as a function of temperature
is shown in Fig.\ \ref{f2}.  The shapes of the curves are
similar to what we found in previous work\cite{AtkinsonII}
in which we examined a model with two planes per unit cell.
Experimental observations
of $\lambda_c$ in Y-123\cite{Anlage,Homes,Munzel} and
Y-124\cite{BasovII}
are contradictory.  All of the experiments find that at low temperatures
$\lambda_c^{-2}(T)$ can be fitted by a linear $T$ dependence,
$\lambda_c(0)^{2}/\lambda_c(T)^2 \sim 1-\alpha T/T_c$, but the
slope of the fit varies dramatically.
Two of the infrared experiments\cite{BasovII,Homes} find that $\alpha
\ll 1$, while the third\cite{Munzel} finds that $\alpha \sim 1$ and
the microwave experiment\cite{Anlage} finds that $\alpha \gg 1$.
Until some sort of consensus is achieved, it will be difficult to
say anything about our model.

\section{Conclusions}
\label{con}
It is clear that our proximity effect model cannot describe the
temperature dependence of the
$ab$ anisotropy of the penetration depth which has been observed
experimentally by Zhang {\em et al.} \cite{Bonn}.
Essentially, the problem with
our proximity model is that, unlike the case of the two--plane
model, there are always regions of the
chain Fermi surface on which the gap is small, so that the
temperature scale over which the penetration depth parallel to the chains
varies is much lower than the scale perpendicular to the chains.
The question that needs to be answered, then, is to what extent is
our model representative of proximity models in general.

The common feature of proximity models is that the chains are
intrinsically normal but driven superconducting by their coupling to the
planes.  Where proximity models differ is in the nature of the
chain--plane coupling.  In our model we have made the assumption
that the chain--plane coupling is coherent, so that chain states
are coupled to plane states with the same value of ${\bf k}$.
The amount of mixing between the two states depends on the difference
in energy between them so that, for example, in Fig.\ \ref{f0}
the chain and plane Fermi surfaces are most strongly mixed in
the neighbourhood of their crossing.  In a similar fashion,
the induced gap on the chain is small (of the order
of a few percent of the intrinsic gap in the plane) wherever the
chain and plane Fermi surfaces are far apart.  This is the reason
for the large difference in the temperature dependence of $\lambda_a$
and $\lambda_b$.

One solution to this is to couple the chains and planes incoherently,
so that every state ${\bf k}$ on the planes is coupled
equally to every state ${\bf k^\prime}$ on the chains.
There is some evidence that there is incoherence along the $c$ axis:
Kleiner and M\"uller\cite{Kleiner} have found an intrinsic Josephson effect
in Bi$_2$Sr$_2$CaCu$_2$O$_8$ and, more recently, in underdoped Y-123.
The d.c.\ resistivity of Y-123 in the normal state\cite{Taillefer} shows
semiconducting behaviour in underdoped samples, and the optical
conductivity along the $c$--axis (see, {\em eg.}, Ref.\ \onlinecite{Homes}
and references contained therein)
has a non--metallic response.  For an incoherent model of the type
described above, the induced gap on the chains is proportional to the average
over the Fermi surface of the gap on the planes.  The difficulty with
this model is that, for a $d$ wave order parameter, the induced gap in
the chains will vanish.  It is, in fact, a general feature of $d$
wave order parameters that they do not contribute to incoherent processes
(see, for example, Refs.\ \onlinecite{Radtke} and \onlinecite{AtkinsonIII}).
If on the other hand, we assume that
the order parameter in the planes has an isotropic $s$ wave symmetry,
then the induced gap on the chains will not vanish.  The problem
now, however, is that incoherent coupling does not introduce a
nodal structure into the gap the way coherent coupling does so that
it is difficult to reconcile such a model with a linear
low temperature penetration depth.  For a model with incoherent
chain--plane coupling to successfully describe the low temperature
penetration depth, it would have to have an order parameter
with nodes on the Fermi surface but whose Fermi surface average
was nonzero, and the induced gap in the chains would have to
be of the order of $T_c$ so that the temperature dependence
of $\lambda_a$ and $\lambda_b$ would be similar.

Leaving, for a moment, the discussion of the nature of the chain--plane
coupling, we will now turn to a more conceptual problem---that
of the size of the chain--plane coupling.  The coupling strength $t_0$
is chosen to account both for the fact that $\lambda_c(T=0)
/\lambda_{a}(T=0) \sim 10$ \cite{BasovII,Anlage,Schutzmann,Homes}
and for the size of the induced gap in
the chains.  As is well known, the chain--plane coupling can
degrade $T_c$ substantially.  We find that for $t_0 = 50 \mbox{ meV}$,
$T_c$ is only 65\% of its value at $t_0 = 0$.
It is also difficult to reconcile the picture of weakly coupled
two dimensional planes with such large values of the chain--plane
coupling.  In our model
the ratio of the electron hopping strengths along the $c$ and
in--plane directions is $t_0/2\sigma_1 = 0.25$, so that it is
difficult to imagine that the $c$ axis coupling is a weak perturbation
in an otherwise two dimensional system.  The challenge, therefore,
for theories which begin with models of a single CuO$_2$ plane
is to explain the large anisotropy between the $a$ and
$b$ supercurrents in YBCO without invoking a large chain--plane coupling.

Kresin and Wolf\cite{Kresin} have suggested that proximity effect
models require an inelastic channel for the chain--plane coupling.
In their two--plane model, electrons can hop between the planes
through coherent tunneling or through scattering from a phonon.
Their model is more three dimensional than the ones discussed above
since the inelastic interplane coupling acts as a pairing process
which leads to an increase in $T_c$.
It is possible that in a chain--plane model, some kind of inelastic
transport mechanism along the $c$ axis might lead to
a sufficiently large gap in the chains that $\lambda_a$ and
$\lambda_b$ would have similar $T$ dependences.  The idea of
a mixture of pairing interactions has recently been proposed
by Song and Annett,\cite{AnnettII} although they have limited
their discussion
to mixing phonons and Coulomb interactions within a single plane.


There is also the issue of whether a simple two--band model can be
representative of Y-123.  More careful band structure calculations
\cite{Pickett,Yu} find that the Fermi surface has four pieces
instead of two.  The two additional pieces of Fermi surface come from
the internal structure of the CuO$_2$ bilayer (which we have treated as
a single layer) and from the internal structure of the CuO chains.
The inclusion of these two pieces of Fermi surface is not likely to
affect the important results contained within this paper however:
the additional piece of Fermi surface due to the CuO$_2$ bilayer has
a nearly tetragonal symmetry (and will therefore contribute to
the anisotropy in the penetration depth) and the piece due to the CuO
chains is small and will only make a small change to the screening
currents.


Our final conclusion, then, is as follows:  A proximity model for
Y-123 in which the superconducting pairing interaction is
localised to the planes and the chain--plane coupling is coherent
will not account for the temperature dependence of the
anisotropy of the penetration depth seen in experiments.  It is
possible that other models for the chain--plane coupling will
be able to adequately describe the $ab$ anisotropy.  The single
largest problem faced by proximity models is that penetration
depth  experiments
\cite{Bonn} seem to indicate that the gap in the chains
is of the same order as the gap in the planes.

\section{Acknowledgments}
This work was supported by a Natural Sciences and Engineering
Research (NSERC) grant, and by the Canadian Institute for Advanced
Research (CIAR).  The authors would like to thank D.\ N.\ Basov
and J.\ S.\ Preston for helpful discussions.

\begin{figure}
\caption{The Fermi surface for the model Hamiltonian is shown for a
range of $k_z$ between $k_z = 0$ and $k_z = \pi/d$.
When $k_z=\pi/d$, the chain--plane coupling vanishes and
the two pieces of Fermi surface are those of the isolated chains and
planes.  As the chain--plane coupling increases, the Fermi surfaces
hybridize and are pushed apart.  The effect of the chain--plane
coupling is largest where the two Fermi surfaces are closest together.
There is an induced gap on the chains whose size is greatest where there
is the most chain--plane mixing.}
\label{f0}
\end{figure}

\begin{figure}
\caption{(a) In--plane penetration depth for an $s$ wave order parameter.
The penetration depth in the $a$ direction (perpendicular to the chains)
is nearly that of a pure $s$ wave superconductor in the absence of
chains.  The penetration depth in the $b$ direction has a very different
temperature dependence from that in the $a$ direction because
the size of the induced gap in the chains is much different from the
size of the gap in the planes.  The relative bandwidths of the
chains and planes were determined by setting
$\lambda_a^2(0)/\lambda_b^2(0) \sim 2.5$, in accordance with
experiment. (b) In--plane penetration depth for a $d$ wave order parameter.
Again, $\lambda_a(T)$ is essentially the same as for a single layer $d$ wave
superconductor, while the shape of $\lambda_b(T)$ reflects the structure
of the induced gap in the chains as well as the planes.}
\label{f1}
\end{figure}

\begin{figure}
\caption{Penetration depth in the $c$ direction for both an $s$ wave
(solid line) and a $d$ wave (dashed line) order parameter.  The
strength of the chain--plane coupling is chosen to be $t_0 = 50 \mbox{ meV}$
so that $\lambda_c^2(0)/\lambda_a^2(0) \sim 100$, as observed experimentally.}
\label{f2}
\end{figure}


\begin{thebibliography}{ambegoakar}

\bibitem{Abrikosov} A. A. Abrikosov, Physica C {\bf 182}, 191 (1991);
A. A. Abrikosov, R.A. Klemm, Physica C {\bf 191}, 224 (1992).

\bibitem{Friedmann} T. A. Friedmann {\em et al.}, Phys. Rev. B {\bf 42},
6217 (1990).

\bibitem{Taillefer} R. Gagnon, C. Lupien and L. Taillefer,
Phys. Rev. B {\bf 50}, 3458 (1994).

\bibitem{Basov} D.  N. Basov {\em et al.}, Phys. Rev. Lett. {\bf 74},
598 (1995).

\bibitem{Bonn} K. Zhang {\em et al.}, Phys. Rev. Lett. {\bf 73}, 2484 (1994).

\bibitem{Tallon} J. L. Tallon {\em et al.}, Phys. Rev. Lett. {\bf 74},
1008 (1995).

\bibitem{Tachiki} M. Tachiki, S. Takahashi, F. Steglich, H. Adrian,
Z. Phys. B {\bf 80}, 161 (1990); M. Tachiki, T. Koyama, S. Takahashi,
Prog. Theor. Phys. Suppl. {\bf 108}, 297 (1992); S. Takahashi,
M. Tachiki, Physica C {\bf 170}, 505 (1990).

\bibitem{Buzdin} A.I. Buzdin, V.P. Damjanovi\'c, A.Yu. Simonov, Phys. Rev. B
{\bf 45}, 7499 (1992); A.I. Buzdin, V.P. Damjanovi\'c, A.Yu. Simonov,
Physica C {\bf 194}, 109 (1992).

\bibitem{Bulaevskii} L. N. Bulaevskii and M. V. Zyskin, Phys. Rev. B
{\bf 42}, 10230 (1990).

\bibitem{Simonov} A. Yu. Simonov, Physica C {\bf 211}, 455 (1993).

\bibitem{KlemmI} R. A. Klemm and S. H. Liu, Phys. Rev. Lett. {\bf 74},
2343 (1995).

\bibitem{Kresin} V. Z. Kresin,
H. Morawitz and S. A. Wolf, {\em Mechanisms of Conventional and High $T_c$
Superconductivity}, Oxford University Press, Oxford (1993); V. Z. Kresin
and S. A. Wolf, Phys. Rev. B {\bf 51}, 1229 (1995).

\bibitem{AtkinsonI} W. A. Atkinson and J. P. Carbotte, Phys. Rev. B
{\bf 51}, 1161 (1995)

\bibitem{AtkinsonII} W. A. Atkinson and J. P. Carbotte, Phys. Rev. B
{\bf 51}, 16371 (1995).

\bibitem{AtkinsonIII} W. A. Atkinson and J. P. Carbotte,
(In Press, Phys. Rev. B).

\bibitem{KlemmII} S. H. Liu and R. A. Klemm, Phys. Rev. B {\bf 45},
415 (1992); R. A. Klemm and S. H. Liu, Physica C {\bf 191}, 383
(1992); {\em ibid.} {\bf 216}, 293 (1993).

\bibitem{Schneider} T. Schneider, H. De Raedt and M. Frick, Z. Phys. B
{\bf 76}, 3 (1989).

\bibitem{Kettemann} S. Kettemann and K. B. Efetov, Phys. Rev. B
{\bf 46}, 8515 (1992).

\bibitem{Pickett} W. E. Pickett, H. Krakaurer, R. E. Cohen and D. J. Singh,
Science {\bf 255}, 46 (1992); W. E. Pickett, R. E. Cohen and H. Krakaurer,
Phys. Rev. B {\bf 42}, 8764 (1990).

\bibitem{Yu} J. Yu, S. Massida, A. J. Freeman and D. D. Koelling,
Phys. Lett. A {\bf 122}, 203 (1987).

\bibitem{Hirsch} J. E. Hirsch and F. Marsiglio, Phys. Rev. B {\bf 45},
4807 (1992).

\bibitem{Mahan} G. D. Mahan, {\em Many-Particle Physics} (Plenum Press,
New York, 1986), p. 30.

\bibitem{Rickayzen} G. Rickayzen, {\em Green's Functions and Condensed
Matter} (Academic Press, San Diego, 1980), p. 8.

\bibitem{BasovII} D. N. Basov, T. Timusk, B. Dabrowski, J. D. Jorgensen,
Phys. Rev. B {\bf 49}, 3511 (1994).

\bibitem{Anlage} J. Mao, D. H. Wu, J. L. Peng, R. L. Greene and S. M.
Anlage, Phys. Rev. B {\bf 51}, 3316 (1995).

\bibitem{Schutzmann} J. Sch\"utzmann, S. Tajima, S. Miyamoto and S. Tanaka,
Phys. Rev. Lett. {\bf 73}, 174 (1994).

\bibitem{Homes}
C. C. Homes, T. Timusk, D. A. Bonn, R. Liang and W. N. Hardy (Submitted
to Physica C).

\bibitem{Munzel} J. M\"unzel {\em et al.}, Physica C {\bf 235-240},
1087 (1994).

\bibitem{Hardy} W. N. Hardy {\em et al.}, Phys. Rev. Lett. {\bf 70},
3999 (1993).

\bibitem{Annett} J. Annett, N. Goldenfeld, S. R. Renn, Phys. Rev. B
{\bf 43}, 2778 (1991).

\bibitem{Prohammer} M. Prohammer, J. P. Carbotte, Phys. Rev. B
{\bf 43}, 5370 (1991).

\bibitem{Kleiner} R. Kleiner, P.  M\"uller, Phys. Rev. B {\bf 49}, 1327
(1994); P. M\"uller, Bulletin of the American Physical Society {\bf 40},
376 (1995).

\bibitem{Radtke} R. J. Radtke, C. N. Lau, K. Levin (Submitted to
Phys. Rev. B).

\bibitem{AnnettII} J. Song and J. F. Annett, Phys. Rev. B {\bf 51},
3840 (1995).

\end{thebibliography}
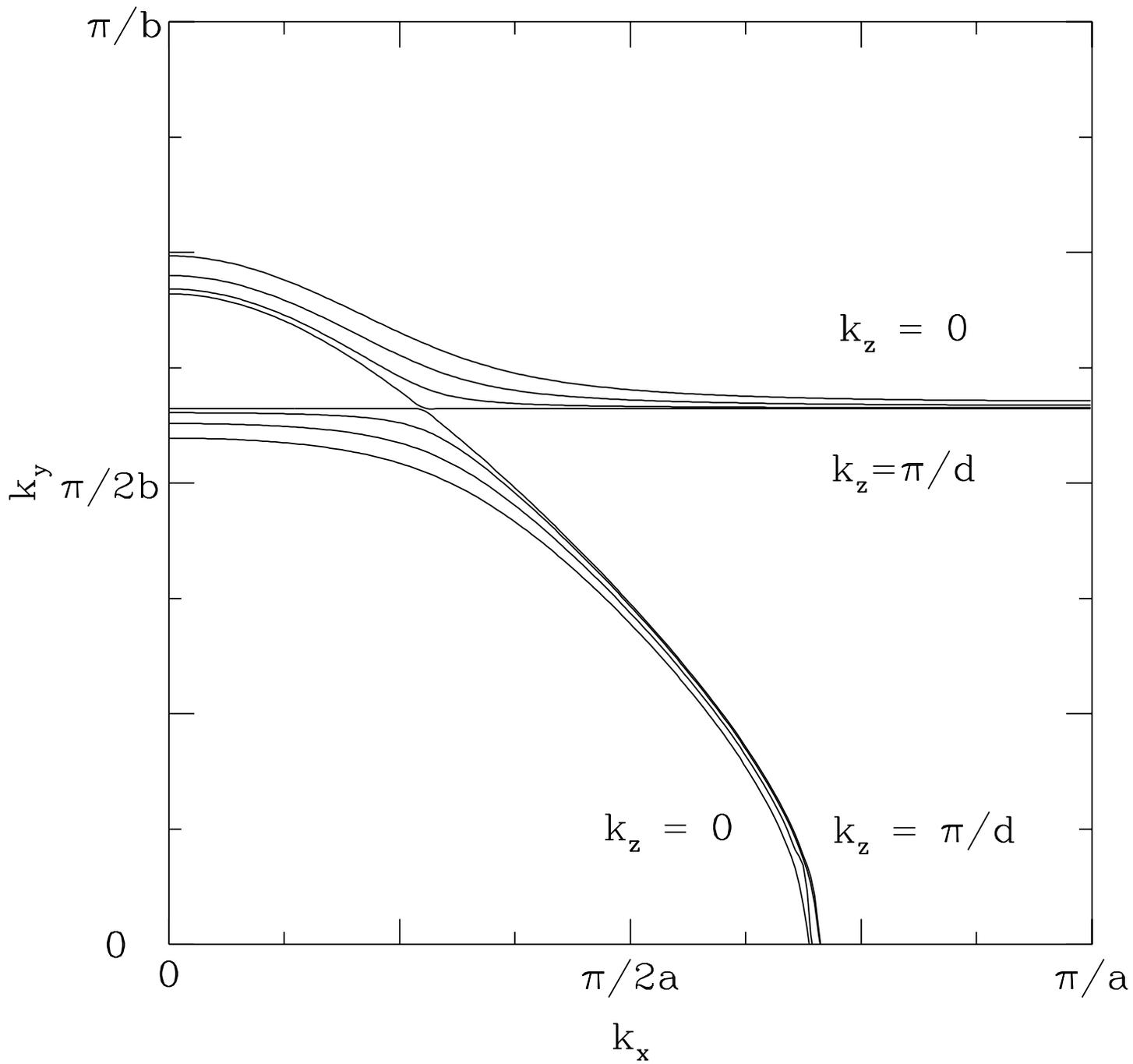
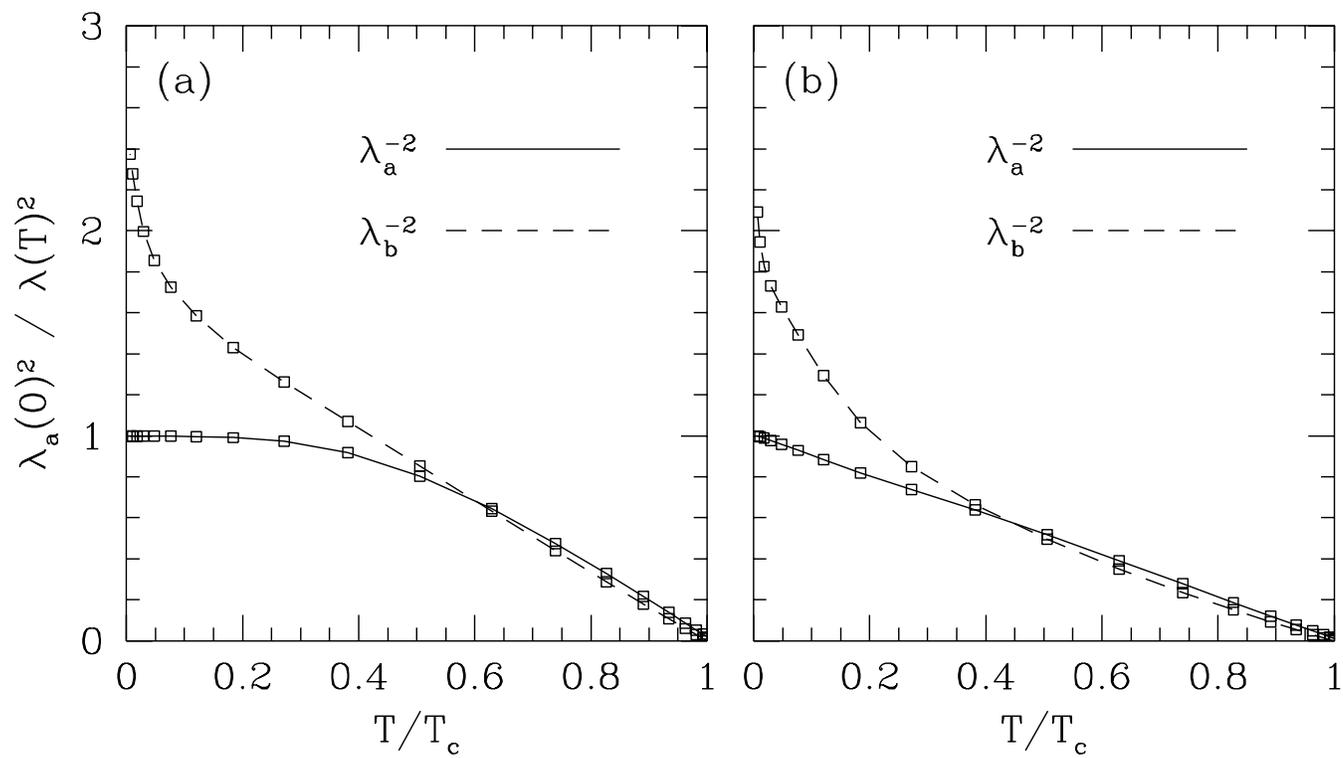
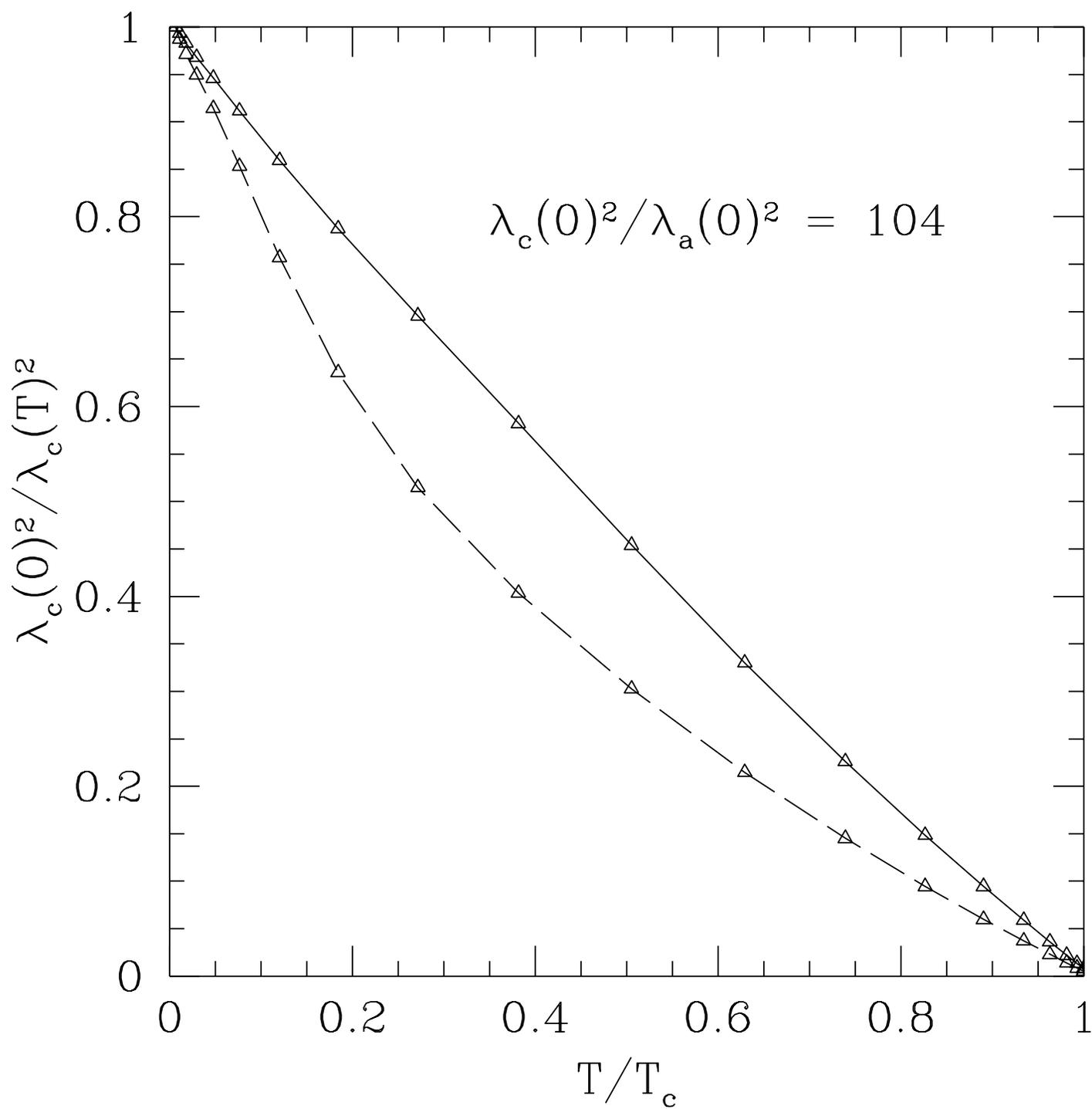
\end{document}